\documentclass{JHEP}        
\usepackage{epsfig}
\preprint{IFT-UAM/CSIC-00-26}

\def\be{\begin{equation}}
\def\ee{\end{equation}}
\def\bea{\begin{eqnarray}}
\def\eea{\end{eqnarray}} 
\def\nn{\nonumber \\}

\def\part{\partial}
\def\tfrac#1#2{{\textstyle{#1\over #2}}}                                  
\def\Tr{\mbox{Tr}}
\def\tfrac#1#2{{\textstyle{#1\over #2}}}
\def\half{\tfrac{1}{2}}

\def\back{\overleftarrow{\mbox{\i}}}
\def\incl{\mbox{i}}

\def\hA{\hat A} 
\def\hC{\hat C} 
\def\hD{\hat D}
\def\hF{\hat F} 
\def\hPhi{\hat{\Phi}}          

\conference{IVth TMR meeting, Paris 2000}

\title{A non-Abelian Chern-Simons term for non-BPS D-branes}

\author{Bert Janssen and Patrick Meessen\\
        Instituto de F{\'\i}sica Te\'orica \\
        Universidad Aut\'onoma de Madrid \\
        C-XVI, C.U. Cantoblanco \\
        28049 Madrid, Spain \\
        E-mail: \email{bert.janssen@uam.es},  
                \email{patrick.meessen@uam.es}}

\abstract{We propose a Chern-Simons term for $N$ coinciding non-BPS D-branes.
Demanding full $U(N)$ invariance and compatibility with T-duality, it is shown 
that it is necessary to introduce new interaction terms, through which the 
non-BPS D-branes couple to all $p$-form RR fields.}

\begin{document}

\section{Introduction}

Non-BPS branes have recently attracted a lot of attention, since they 
might give us insight in the behaviour of $p$-branes without the protection of
supersymmetry.    
Non-BPS branes have (open) tachyonic modes living on their world volume, due 
to the lack of supersymmetry, which make them in general unstable objects. 
However, when the tachyon potential has minima, very interesting 
phenomena can occur. It has been conjectured by Sen \cite{Sen} that at the 
stationary point of the potential, the negative energy of the potential and 
the positive tension of the brane cancel each other, creating a state which is 
indistinguishable from the vacuum. 
Even more interesting things happen when the tachyon field interpolates between
two minima of the potential: then, after condensation of the tachyon living on
a (unstable) non-BPS-brane into a kink solution, the non-BPS brane reduces to 
a (stable) BPS $(p-1)$-brane \cite{Sen1}-\cite{Hor}.

The dynamics of these non-BPS branes is given by their Born-Infeld (BI) and 
Chern-Si\-mons (CS) action, including a tachyon field coupling, which 
distinguishes it from the actions of ordinary BPS branes. These actions have 
been recently given in \cite{BCR}-\cite{Kluson2}. 

Another interesting topic that lately has become popular, is the phenomenon of
gauge symmetry enhancement if we let $N$ (BPS) D-branes coincide \cite{Witten}.
As the $N$ D-branes approach each other, the open strings stretched between 
them become massless, the $N$ $U(1)$ BI vectors are promoted to a $U(N)$ 
valued gauge field $A^a$
and the $(9-p)$ scalars $\Phi^i$ that describe the position of the branes in 
the transverse directions become scalars in the adjoint representations of 
$U(N)$, gene\-ra\-ting a non-com\-mutative geometry in the transverse space. 

The (non-commutative) BI and CS term for coinciding D-branes was 
recently given by \cite{Myers}. There it was argued by means of T-duality 
that some new terms should appear due to the non-Abelian character of the 
sca\-lars $\Phi^i$. These new terms give rise to a non-trivial potential for 
the sca\-lars, cau\-sing a kind of di\-elec\-tric ef\-fect on the D$p$-branes 
in the presence of higher form-fields \cite{Myers}.

The aim of this letter is to unify the results of \cite{BCR}-\cite{Kluson} and 
\cite{Myers}, constructing a non-commutative CS term for non-BPS 
D-branes.\footnote{A proposal for a non-commutative BI term for non-BPS 
D-branes was given recently in \cite{Garousi}.} In \cite{BCR}-\cite{Kluson}, 
there already have been given different proposals for CS terms for $N$ 
coinciding D-branes. However, since \cite{Myers}, it has become clear that 
extra couplings and extra terms have to be taken into account to guarantee the 
$U(N)$ invariance of the action and the correct behaviour under T-duality. In 
this letter we will try to give a more complete action, which takes these 
features into account. 

The organisation of this letter is as follows: in the next two sections we 
will briefly review the CS term for coinciding (BPS) D-branes and  for 
non-BPS branes and  point out the subtleties occurring when trying to unify 
these. In section 4, we will present our proposal for the CS term for $N$ 
coinciding non-BPS D-branes and show that it is compatible with $T$-duality.

For simplicity, throughout this letter we will work in flat space-time with the
Kalb-Ramond field $B_{\mu\nu}$ set equal to zero (i.e. neglecting all effects 
coming from the off-diagonal parts of the metric and its mixing with the 
Kalb-Ramond form in the T-duality rules). Later on in this letter, we will 
propose in analogy with \cite{Myers},  what we believe is the full CS 
term in the presence of $B$.   

\section{A non-commutative CS term for BPS branes}
\label{chapter}
It is well known that BPS D$p$-branes couple to the $(p+1)$-form RR 
fields \cite{Polch, GHM} through their CS term in the following 
way:\footnote{Modulo terms involving the A-roof hat genus, which we will 
neglect everywhere in this letter.}

\be
S= \int_{p+1} C_{n} \wedge e^F \ .
\label{GHMaction}
\ee
Here $F$ is  the field strength of the $U(1)$ BI vector living in the world
volume of the D-brane and with $C_{n}$ we denote the pull-back of the (formal 
sum over) $n$-form RR fields, e.g.
\bea
C_{a_1a_2} &=& 
       C_{\mu_1\mu_2}\  \part_{a_1} X^{\mu_1}\part_{a_2} X^{\mu_2} \ ,
        \label{pullback}      \nn
      &=& C_{a_1a_2} -2 C_{i[a_1} \ \part_{a_2]} \Phi^{i} 
            + C_{i_1i_2} \ \part_{a_1} \Phi^{i_1}\part_{a_2} \Phi^{i_2}
           \nonumber 
\eea
where we have split the space-time coordinates $X^\mu$ in world volume 
coordinates $\xi^a$ and transverse coordinates $\Phi^i$.

It was pointed out in \cite{Witten} that a remarkable gauge symmetry 
enhancement occurs when $N$ parallel branes approach each other: the ground
states of the strings stretched between the various D-branes become massless, 
the $N$ Abelian BI vectors become the components of a non-Abelian $U(N)$ 
vector and the scalars $\Phi^i$, indicating the position of the branes in the 
transverse direction, rearrange into a set of non-Abelian scalars, transforming
in the adjoint representation of $U(N)$. In other words, the $U(1)^N$ symmetry 
group of the $N$ D-branes is enhanced to a $U(N)$ gauge group.

Recently it was shown how the CS term (\ref{GHMaction}) gets modified under 
this effect \cite{Myers}. Of course, one obvious change is the necessity to 
take the trace over the $U(N)$ indices. But there is more: first of all, it 
turns out that the background fields should be functions of the 
non-commuta\-ti\-ve 
sca\-lars $\Phi^i$. Secondly, since the pull-back with respect to the 
transverse coordinates $\Phi^i$ has become $U(N)$-valued, the partial 
de\-ri\-va\-tives of $\Phi^i$ in (\ref{pullback}) have to be replaced by 
covariant de\-ri\-va\-tives:
\be
D \Phi^i = \part \Phi^i + i g [A, \Phi^i] \ , 
\ee
where $A$ is the $U(N)$ gauge field with its field strength tensor given by
\be
F= dA + \tfrac{ig}{2} [A, A] \ .
\ee
And finally it turns out that T-duality requires some extra interaction terms,
due to the non-Abelian character of the scalars $\Phi^i$.

The CS term (\ref{GHMaction}) is compatible with T-duali\-ty, in the sense that
a T-duality transformation maps the term of a D$p$-brane into the term of a
D$(p\pm1)$-brane \cite{ABB, BR}. This is easily shown by mapping the CS term 
of the D$p$ and D$(p-1)$-brane, via double and direct dimensional reduction 
respectively, onto the same term in nine dimensions.\footnote{This technique
is the same as the one used to show the T-duality between the RR fields in  
Type IIA and  B supergravity \cite{BHO, MO}.} Double dimensional reduction
of the D$p$-brane term gives:\footnote{Performing dimensional reduction, it is 
necessary to distinguish between ten- and nine-dimensional fields. We will do 
this by indicating the ten-dimensional ones with a hat. However in the parts 
where there is no confusion, the hats will be omitted.}
\be
S = \int_{p+1} \hspace{-.4cm}  \hC^+ \wedge e^{\hF}  
  = \int_{p}  ( C^- + C^+ d \chi )\wedge e^{F}   \ ,
\label{p}
\ee
where the upper index $+$ denotes the even (odd) RR-forms and the $-$ the odd 
(even) RR-forms. The nine-dimensional scalar field $\chi$, from the  
ten-dimensional point of view, comes from the component of 
the BI vector in the direction over which we reduce: $\chi = \hA_{p}$. 
The nine-dimensional term is exactly the one obtained after 
direct dimensional reduction of the D$(p-1)$-brane term
\be
S = \int_{p} \hC^- \wedge e^{F}  
  = \int_{p}  ( C^- + C^+ d \chi )\wedge e^{F}  \ ,
\label{p-1}
\ee
but now the scalar $\chi$ comes from the reduction over the transverse 
direction: $\chi =\hPhi^{p}$. Note that in (\ref{p-1}),  $F$ is a 
world volume field therefore remains invariant under direct reduction. Thus 
the $p$-th 
component of the BI vector $\hA_{p}$ gets mapped into the extra transverse 
direction $\hPhi^{p}$ (or vice versa) and hence the $\hF_{ap}$ component of 
the field strength into the pull-back $\part_a \hPhi^{p}$. 

However, after the symmetry enhancement, non-trivial commutators appear in the
pull-backs and field strengths and the rules for (double) dimensional reduction
become more involved. The double dimensional reduction of the D$p$-brane term
now gives (compare to (\ref{p})):\footnote{By Tr we mean the symmetric trace
description of \cite{Tseyt, Tseyt2, Myers}.}
\bea
S &=& \int_{p+1} \Tr \{ \hC^+ \wedge e^{\hF} \} 
           \label{ncp-1} \\
  & =& \int_{p} \Tr \{( C^- + C^+ d \chi  +  ig[\chi, \Phi^i] C^+_i)
                         \wedge e^{F} \} \ ,
\nonumber
\eea
where the last term now comes from the reduction of the non-Abelian pull-back
\be
\hD_p \hPhi^i = i g [\chi,\Phi^i] \ . 
\label{redphi}
\ee
Clearly, an extra term has to be added to the term (\ref{p-1}) in order to
reduce to the same nine-dimensional term (\ref{ncp-1}). The term
\be
S = \int_{p} \Tr \{(\hC^- + ig\ \incl_{\hPhi} \incl_{\hPhi} \hC^- ) 
              \wedge e^{F} \} 
\label{order1}
\ee 
does reduce in the correct way, where  
\bea
\left(\incl_\Phi\incl_\Phi C^{(n)}\right)_{a_1...a_{n-2}}
     &\equiv&\Phi^j \Phi^i C_{ija_1...a_{n-2}}
\label{incl} 
       = \half [\Phi^j, \Phi^i] C_{ija_1...a_{n-2}} 
\nonumber .
\eea 
Applying again T-duality on (\ref{order1}) shows the need to include a new 
term of the form
\be
( ig\ \incl_{\hPhi}\incl_{\hPhi})^2 \hC^+
\ee 
and an iterative procedure gives the fully T-dua\-li\-ty invariant CS term:
\be
S= \int_{p+1} \Tr\{ e^{ig\incl_\Phi\incl_\Phi} C \wedge e^{F} \} \ .
\label{Myersaction}
\ee  
Thus we see that the $N$ coinciding D$p$-branes not only couple to the 
$(p+1)$-form RR fields and the lower ($p-1$, $p-3$, ...) RR-forms via the 
$F$ as in the Abelian case, but also to higher ($p+3$, $p+5$, ...) RR-forms 
via the the extra contraction terms (\ref{incl}). It is these contraction 
terms that give rise to multipole moments and lead to a kind of polarisation 
effect of D-branes in the present of the external RR field \cite{Myers}. 

\section{Non-BPS branes and their proposed CS term}

Unlike the well-known supersymmetric D-branes, a (single) non-BPS D$p$-brane 
couples to the $p$-form RR field \cite{Hor} and therefore even (odd) non-BPS 
D$p$-branes appear in Type IIB (IIA) string theory. 

In \cite{Sen3}-\cite{BCR}, a CS term was suggested that ge\-nerates this 
coupling:
\be
S = \int_{p+1} C_n \wedge dT \wedge e^{F} \ .
\label{BCRaction} 
\ee
The form of this term was argued from string scattering amplitudes
and the fact that after ta\-chyon condensation, (\ref{BCRaction}) reduces 
correctly to the well-known CS term for BPS D-branes \cite{GHM}. The authors 
of \cite{BCR} also give a generalisation to the non-Abelian case of $N$ 
coinciding D-branes, by taking the trace over the indices of the $U(N)$ 
symmetry group:
\be
S = \int_{p+1} C_n \wedge d\ \Tr\{ T \wedge e^{F}\} \ ,
\label{BCR2action} 
\ee
where $T$ is in the adjoint representation of $U(N)$.

In \cite{KW} it was shown that for a system of $N$ coinciding D-branes 
and anti-D-branes, an interaction term with higher powers of the tachyon is 
needed in the CS term. This leads, before tachyon condensation, to the 
gene\-ra\-lisation of the term (\ref{BCR2action}) to \cite{Kluson}:
\bea
S&=&\sum_{k,l} a_{kl} S_{kl} \ , 
\label{KWKaction}\\
S_{kl} &=& \int_{p+1} C_n \wedge 
            \Tr \Bigl\{ (DT)^{2k+1} T^{2l} \wedge e^{F}\Bigr\} \ ,
\nonumber
\eea
where the coefficients $a_{kl}$ are undetermined numerical con\-stants and
\be
D T = d T + i g [A, T] \ . 
\ee
 It was shown that this term reproduces correctly the CS term for BPS 
D-branes after tachyon condensation. 

Yet one can try to go a step further in the attempt to construct
a non-Abelian CS term and demand invariance under T-duality.
As shown in section \ref{chapter}, the pull-back of the bulk fields have to 
be covariantised and extra interaction terms have to be included in order to 
have a correct behaviour under T-duality. In the next section we will try to 
construct such an term.

\section{A fully $U(N)$-invariant CS term for non-BPS D-branes}

Our proposal for the CS term of $N$ coinciding non-BPS D$p$-branes is
\bea
S&=& \sum_{kl} a_{kl}\int_{p+1} 
            \Tr   \Bigl \{ e^{ig\incl_\Phi\incl_\Phi} C 
                         \Bigl[ -ig \back_{[\Phi, T]} + DT\Bigr] 
                                 \times \ (DT)^{2k} T^{2l} e^F  \Bigr \} \ ,
\label{onze}
\eea 
where 
\be
\Bigl(C^{(n)} \back_{[\Phi, T]}\Bigr)_{a_1...a_{n-1}} = C_{a_1...a_{n-1}i} [\Phi^i, T] \ . 
\label{back}
\ee
Like in section \ref{chapter}, we will demonstrate the correct behaviour 
under T-duality through 
double and single dimensional reduction. Let us first consider the case where
$k=l=0$, which corresponds to a generalisation of the term (\ref{BCR2action}):
\bea
~\hspace{-.5cm}  
S= \hspace{-.1cm}        \int_{p+1} \hspace{-.4cm} \Tr 
                 \Bigl \{ e^{ig\incl_{\hPhi}\incl_{\hPhi}} \hC 
                         \Bigl[ -ig \back_{[\hPhi, T]} + \hD T\Bigr]  e^{\hF} 
                  \Bigr \}.
\label{onzeBCR}
\eea  
Double dimensional reduction of the second term  gives
\bea
&&\hspace{-.4cm} 
e^{ig\incl_{\hPhi}\incl_{\hPhi}} \hC^+ \hD T  e^{\hF} =\Bigl\{
      - e^{ig\incl_{\Phi}\incl_{\Phi}} C^- D T 
       +\ ig \ e^{ig\incl_{\Phi}\incl_{\Phi}} C^+ \back_{[\Phi, \chi]}  D T\nn
&& \hspace{3.3cm}  
   + \ ig \ e^{ig\incl_{\Phi}\incl_{\Phi}} C^+ [\chi, T] 
     \  - \  e^{ig\incl_{\Phi}\incl_{\Phi}} C^+ D\chi DT 
\Bigl\} \ e^F\ ,\label{term1}
\eea
where again $\chi = \hA_p$. The reduction of the first term gives
\bea
&&
e^{ig\incl_{\hPhi}\incl_{\hPhi}} \hC^+  \back_{[\hPhi, T]} e^{\hF} = 
   \Bigl\{ 
        - \ e^{ig\incl_{\Phi}\incl_{\Phi}} C^- \back_{[\Phi, T]}
 \  + \ e^{ig\incl_{\Phi}\incl_{\Phi}} C^+ \back_{[\Phi, T]} D\chi
\nn
&&\hspace{4.2cm} 
    - ig \ e^{ig\incl_{\Phi}\incl_{\Phi}} C^+ \back_{[\Phi, T]}
                           \back_{[\Phi, \chi]}  
               \Bigr\}\  e^F \ .
\label{term2}
\eea
On the other hand, the single reduction of the D$(p-1)$-brane CS term yields 
the terms
\bea
e^{ig\incl_{\hPhi}\incl_{\hPhi}} \hC^- &=&
            e^{ig\incl_{\Phi}\incl_{\Phi}} C^- 
       \  +\  e^{ig\incl_{\Phi}\incl_{\Phi}} C^+ D\chi 
       \  - \ ig \ e^{ig\incl_{\Phi}\incl_{\Phi}} C^+ \back_{[\Phi, \chi]}  \ ,
\label{termI}
\\
e^{ig\incl_{\hPhi}\incl_{\hPhi}} \hC^- \back_{[\hPhi, T]} &=&
          e^{ig\incl_{\Phi}\incl_{\Phi}} C^- \back_{[\Phi, T]} 
      \ - \ e^{ig\incl_{\Phi}\incl_{\Phi}} C^+ \back_{[\Phi, T]} D\chi \nn
  &&  \ +\ ig \ e^{ig\incl_{\Phi}\incl_{\Phi}} C^+ \back_{[\Phi, T]} 
                               \back_{[\Phi, \chi]}
      \ -\ ig \  e^{ig\incl_{\Phi}\incl_{\Phi}} C^+ [ \chi, T] \ . 
\label{termII}
\eea
Given that the world volume fields $DT$ and $F$ do not change under single 
reduction, we see that the terms (\ref{term1})-(\ref{term2}) obtained from the 
double reduction of the D$p$-brane CS, coincide with the terms 
(\ref{termI})-(\ref{termII}) coming from the single reduction of the 
D$(p-1)$-brane. This proves the T-duality invariance of (\ref{onzeBCR}).

Comparing (\ref{term1}) and (\ref{termII}), it is clear why the extra term  
\be
e^{ig\incl_{\hPhi}\incl_{\hPhi}} \hC^+  \back_{[\hPhi, T]}
\ee 
has been introduced: double reduction of the second term in (\ref{onzeBCR})
yields a term proportional to the commutator $[\chi, T]$, as a result of the
reduction of $\hD T$ (analogous to (\ref{redphi})) and this term can only be 
compensated by a term of the form (\ref{termII}). 

The generalisation of the proof to $k,l \neq 0$ is straightforward: the only 
extra contributions come from double reduction the $(\hD T)^{2k}$ terms. It is 
easy to see that
\bea
(\hD T)^n &=& (DT + ig\ [\chi, T] d\xi^p )^n \nn
          &=& (DT)^{n-1} 
               \Bigl( DT+ ig\  [\chi, T] \delta_{n, 2k+1} d\xi^p \Bigr) \ ,
\label{DTn}
\eea 
where the commutator term is only different from zero for odd power of 
$\hD T$, due to the anti-sym\-me\-tric character of the wedge product and the 
symmetric trace prescription. 
Hence the contribution of (\ref{DTn}) factorizes into a term already 
present in the $k,l=0$ case and an extra overall $(DT)^{2k}$. This completes 
the prove for the T-duality invariance of our proposed CS term (\ref{onze}).

Finally we will extend the CS term (\ref{onze}) in the presence of the 
Kalb-Ramond field $B$. In analogy with \cite{Myers}, we propose 
\bea
S \hspace{-.1cm} 
&=&\hspace{-.1cm}  \sum_{kl} a
_{kl} \hspace{-.1cm} 
        \int_{p+1}
              \Tr  \Bigl \{ e^{ig\incl_\Phi\incl_\Phi}( C e^B) 
                         \Bigl[ -ig \back_{[\Phi, T]} + DT\Bigr]  
                                 \times \ (DT)^{2k} T^{2l} e^F  \Bigr \} \ .
\label{onzeB}
\eea 

It would be interesting to see whether the proposed action, in combination 
with the BI term of \cite{Garousi}, contains solutions with dynamical tachyon
condensation and a kind of polarisation effect on the non-BPS branes.    
We leave this for further investigation.

\acknowledgments
The authors thank C. G\'omez, T. Ort{\'\i}n and P. Silva for the useful 
discussions. This work has been partially supported by the TMR program 
FMRX-CT96-0012 on 
{\sl Integrability, non-perturbative effects, and symmetry in quantum field 
theory}.


\end{document}